\documentclass[a4paper,11pt]{article}
\usepackage{pos}

\title{The cosipy library: COSI’s high-level analysis software}
 \ShortTitle{The cosipy library}

\author*[u,f]{Israel Martinez-Castellanos}

\author[r]{Savitri Gallego}
\author[q]{Chien-You Huang}
\author[f]{Chris Karwin}
\author[f]{Carolyn Kierans}
\author[r]{Jan Peter Lommler}
\author[h]{Saurabh Mittal}
\author[d]{Michela Negro}
\author[f,l]{Eliza Neights}
\author[b]{Sean N.  Pike}
\author[c]{Yong Sheng}
\author[h]{Thomas Siegert}
\author[h]{Hiroki Yoneda}
\author[a]{Andreas Zoglauer}

\author[a]{John A. Tomsick}
\author[b,a]{Steven E. Boggs}
\author[c]{Dieter Hartmann}
\author[c]{Marco Ajello}
\author[d]{Eric Burns}
\author[e]{Chris Fryer}
\author[a]{Alexander Lowell}
\author[g]{Julien Malzac}
\author[b]{Jarred Roberts}
\author[a]{Pascal Saint-Hilaire}
\author[f]{Albert Shih}
\author[i]{Clio Sleator}
\author[j]{Tadayuki Takahashi}
\author[k]{Fabrizio Tavecchio}
\author[i]{Eric Wulf}
\author[a]{Jacqueline Beechert}
\author[a]{Hannah Gulick}
\author[a]{Alyson Joens}
\author[a]{Hadar Lazar}
\author[a]{Juan Carlos Martinez Oliveros}
\author[j]{Shigeki Matsumoto}
\author[j]{Tom Melia}
\author[m]{Mark Amman}
\author[c]{Dhruv Bal}
\author[g]{Peter von Ballmoos}
\author[c]{Hugh Bates}
\author[n]{Markus B\"ottcher}
\author[o]{Andrea Bulgarelli}
\author[p]{Elisabetta Cavazzuti}
\author[q]{Hsiang-Kuang Chang}
\author[a]{Claire Chen}
\author[q]{Che-Yen Chu}
\author[o]{Alex Ciabattoni}
\author[p]{Luigi Costamante}
\author[n]{Lente Dreyer}
\author[o]{Valentina Fioretti}
\author[p]{Francesco Fenu}
\author[k]{Giancarlo Ghirlanda}
\author[i]{Eric Grove}
\author[g]{Pierre Jean}
\author[c]{Nikita Khatiya}
\author[g]{J\"urgen Kn\"odlseder}
\author[s]{Martin Krause}
\author[c]{Mark Leising}
\author[f,w]{Tiffany R. Lewis}
\author[t]{Lea Marcotulli}
\author[a]{Samer Al Nussirat}
\author[v]{Kazuhiro Nakazawa}
\author[r]{Uwe Oberlack}
\author[d]{David Palmore}
\author[o]{Gabriele Panebianco}
\author[o]{Nicolo Parmiggiani}
\author[f]{Tyler Parsotan}
\author[a]{Field Rogers}
\author[n]{Hester Schutte}
\author[f]{Alan P. Smale}
\author[i]{Jacob Smith}
\author[d]{Aaron Trigg}
\author[f]{Tonia Venters}
\author[j]{Yu Watanabe}
\author[f]{Haocheng Zhang}

\affiliation[a]{Space Sciences Laboratory, 7 Gauss Way, University of California, Berkeley CA 94720-7450, USA}

\affiliation[b]{Department of Astronomy \& Astrophysics, UC San Diego, 9500 Gilman Drive, La Jolla CA 92093, USA}

\affiliation[c]{Clemson University, South Carolina, USA}

\affiliation[d]{Louisiana State University, Baton Rouge, LA 70803, USA}

\affiliation[e]{Los Alamos National Laboratory, New Mexico, USA}

\affiliation[f]{NASA Goddard Space Flight Center, 8800 Greenbelt Road, Greenbelt, MD 20771, USA}

\affiliation[g]{Institut de Recherche en Astrophysique et Planetologie, 9, avenue du Colonel Roche BP 44346 31028 Toulouse Cedex 4, France}

\affiliation[h]{Institut f{\"u}r Theoretische Physik und Astrophysik, Universit{\"a}t W{\"u}rzburg, Campus Hubland Nord, Emil-Fischer-Str. 31, 97074, W{\"u}rzburg, Germany}

\affiliation[i]{U.S. Naval Research Laboratory, 4555 Overlook Ave., SW Washington, DC 20375, USA}

\affiliation[j]{Kavli Institute for the Physics and Mathematics of the Universe, The University of Tokyo, Japan}

\affiliation[k]{Istituto Nazionale di Astrofisica, Merate, Italy}

\affiliation[l]{George Washington University, USA}

\affiliation[m]{Independent, USA}

\affiliation[n]{Centre for Space Research, North-West University, Potchefstroom 2520, South Africa}

\affiliation[o]{Istituto Nazionale di Astrofisica, Bologna, Italy}

\affiliation[p]{Italian Space Agency, Italy}

\affiliation[q]{Institute of Astronomy, National Tsing Hua University, Guangfu Rd., Hsinchu City, 300044, Taiwan}

\affiliation[r]{Institut f{\"u}r Physik \& Exzellenzcluster PRISMA\textsuperscript{+}, Johannes Gutenberg-Universit{\"a}t Mainz, 55099 Mainz, Germany}

\affiliation[s]{Centre for Astrophysics Research, Department of Physics, Astronomy and Mathematics, University of Hertfordshire, College Lane, Hatfield AL10 9AB, UK}

\affiliation[t]{Yale University, USA}

\affiliation[u]{University of Maryland, USA}

\affiliation[v]{Nagoya University, Japan}

\affiliation[w]{NASA Postdoctoral Program Fellow}

\emailAdd{israel.martinez@nasa.gov}

\abstract{
The Compton Spectrometer and Imager (COSI) is a selected Small Explorer (SMEX) mission launching in 2027. It consists of a large field-of-view Compton telescope that will probe with increased sensitivity the under-explored MeV gamma-ray sky (0.2-5 MeV). We will present the current status of cosipy, a Python library that will perform spectral and polarization fits, image deconvolution, and all high-level analysis tasks required by COSI’s broad science goals: uncovering the origin of the Galactic positrons, mapping the sites of Galactic nucleosynthesis, improving our models of the jet and emission mechanism of gamma-ray bursts (GRBs) and active galactic nuclei (AGNs), and detecting and localizing gravitational wave and neutrino sources. The cosipy library builds on the experience gained during the COSI balloon campaigns and will bring the analysis of data in the Compton regime to a modern open-source likelihood-based code, capable of performing coherent joint fits with other instruments using the Multi-Mission Maximum Likelihood framework (3ML). In this contribution, we will also discuss our plans to receive feedback from the community by having yearly software releases accompanied by publicly-available data challenges.
}

\ConferenceLogo{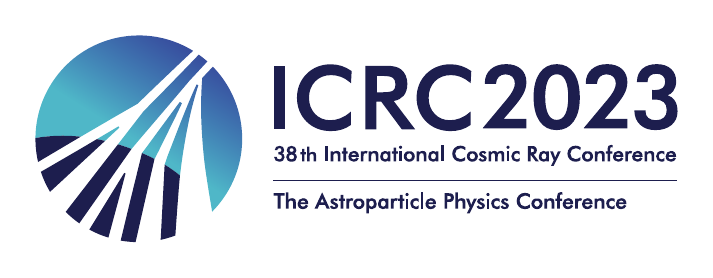}

\FullConference{%
38th International Cosmic Ray Conference (ICRC2023)\\
  26 July - 3 August, 2023\\
  Nagoya, Japan}

\newcommand{\lrp}[1]{\left(#1\right)}

\graphicspath{{./figures}}

\begin{document}
\maketitle

\section{Overview}

The Compton Spectrometer and Imager (COSI)\cite{COSIofficial} is a compact Compton telescope sensitive to gamma rays between 200 keV and 5 MeV.  COSI is a survey instrument with a large field of view ---$> \pi$ sr---   excellent energy resolution  ---$<$1.2\% ($<$0.07\%) at 511 keV (1.157 MeV)--- capable of imaging with a good angular resolution --- $< 4.1^\circ$($<2.1^\circ$) at 511 keV (1.809 MeV)-- and performing polarization measurements.  COSI will reduce the so-called ``MeV gap'',  the low-energy part of the gamma-ray spectrum that despite its great scientific potential is currently under explored.  COSI will uncover the origin of the Galactic positrons,  advance our understanding of how the elements were formed,  and probe the physics of extreme environments and multi-messenger sources. 

The COSI team is currently developing cosipy,  a Python library that will be capable of performing all the high-level task expected for COSI: imaging,  spectral analysis and polarimetry.  On this contribution we first introduce, in Section \ref{sec:basics}, the basic principles of Compton data analysis.  This is then followed by the design of cosipy and its current status.  We also include our plans for the gradual improvement of cosipy and public data challenges, leading up to the expected launch of COSI in 2027.

\section{Compton data analysis}
\label{sec:basics}

\begin{figure}
\centering
\includegraphics[width=.4\textwidth]{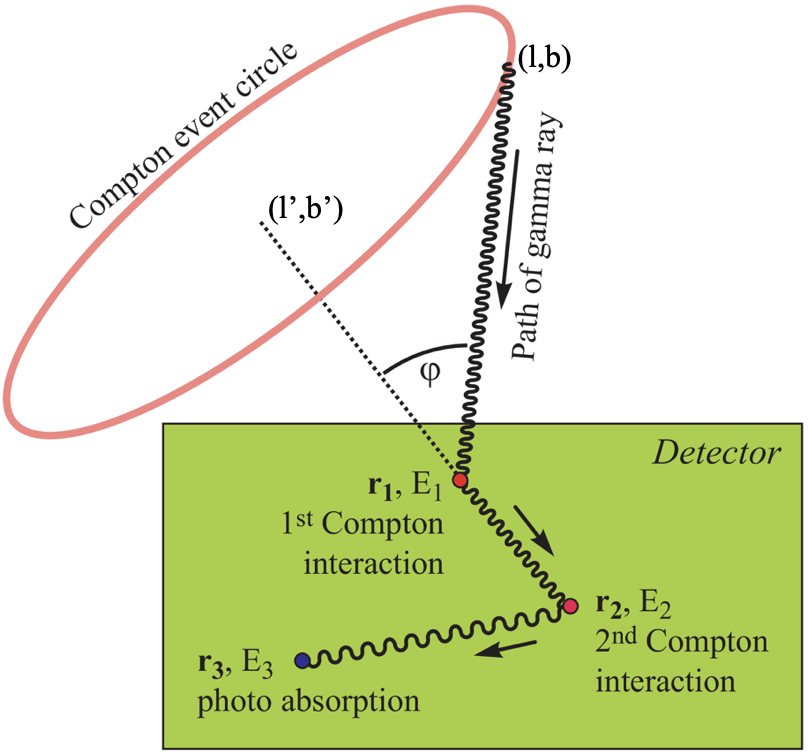}
\caption{COSI measures the location and the energy deposited by the different interactions of a gamma ray with the detector. During reconstruction, this path is deduced based on the kinematics of Compton scattering, constraining the direction of the incoming photon to a circle in the sky, centered in the direction of the first scattered gamma ray (l',b') and with a radius equal to the scatting angle of the first interaction $\phi$.}
\label{fig:reco}
\end{figure}

In the MeV regime photons interact with the detector predominantly through Compton scattering.  As shown in Fig. \ref{fig:reco}, the incoming photon results in a series of energy deposits in the detector until it is finally absorbed due to the photoelectric effect.  Compact telescope like COSI cannot resolve the order of the interactions based on their timing, and therefore it must be derived based on the kinematics of Compton scattering. Once this is determined, the scattering angle $\phi$ and direction (l',b') of the outgoing photon in the first interaction define a circle in the sky that contains the direction of the incoming photon (l,b).  The set of parameters \{$\phi$, l', b'\} is usually called the Compton data space (CDS). Although the following interactions also contain useful information,  the CDS is the minimum information needed to perform imaging and polarimetry.  

\begin{figure}
\centering
\includegraphics[width=.9\textwidth]{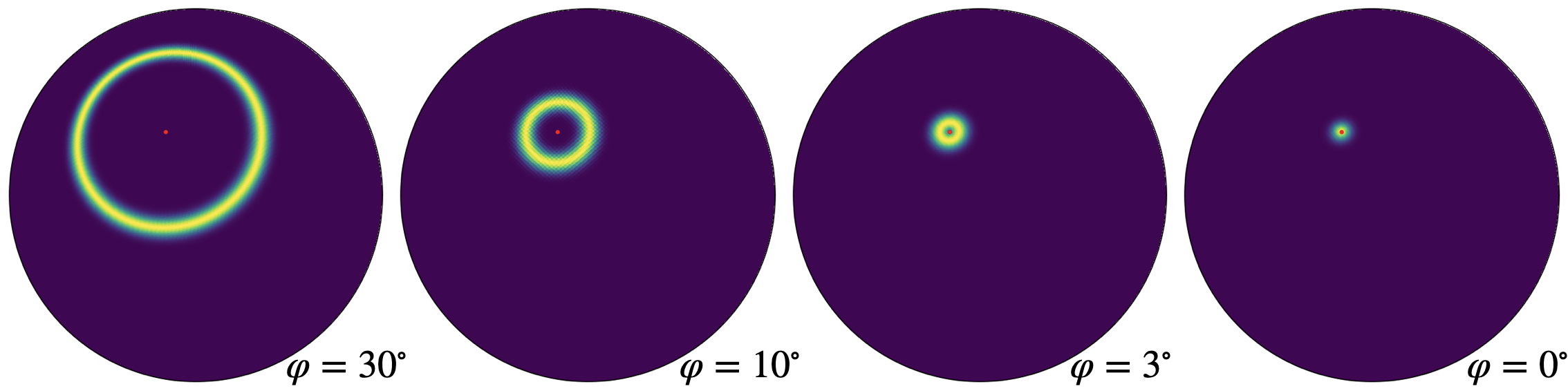}
\caption{Different slices of a simplified point spread function (PSF) in the Compton data space (CDS).  The PSF was sliced along the $\phi$-axis (scattering angle) and shown in an ortographic projection of the sphere describing the direction of the outgoing scattered photon. The red dot shows the true location of the incoming photon.}
\label{fig:compton_cone_slices}
\end{figure}

Since the radius of the Compton circle is fully constrained by the scattering angle $\phi$, the combination over all possible values of $\phi$ define a cone in the three-dimensional CDS.  The walls of this cones are broaden due to instrumental effects, such as the unknown kinetic energy of the electron that participated in the first interaction, the finite energy resolution,  missed interactions due to the detection threshold, etc. The distribution of the broaden Compton cone is the point spread function (PSF) of the instrument, as shown in Fig. \ref{fig:compton_cone_slices}. 

\begin{figure}
\centering
\includegraphics[width=.9\textwidth]{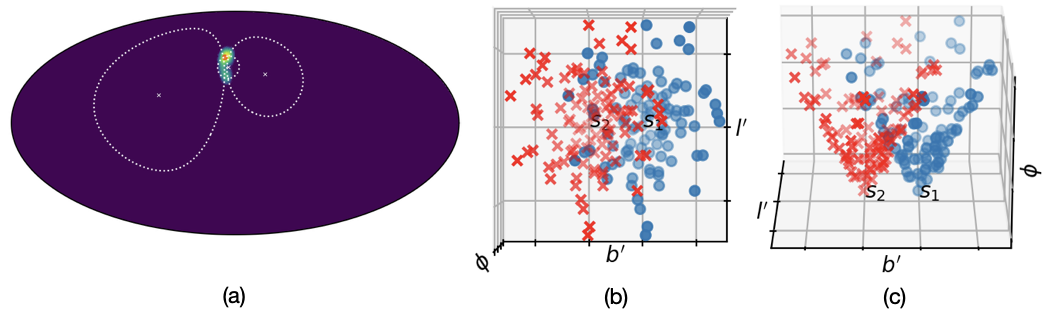}
\caption{(a) A simplified example demonstrating the localization of a source (red dot) based on three detected events (white crosses, with Compton circles shown with dashed lines). (b,c) Events produced by two sources (s$_1$ and s$_2$) distributed in the Compton Data Space (CDS). Views from the top (b) and side (c).  By keeping track of the scattered angle $\phi$ the cones can be discriminated and the image deconvolved.}
\label{fig:localization}
\end{figure}

A source can be localized by combining the information from multiple events,  as shown in Fig \ref{fig:localization}.  The imaging of the sky involves multiple sources and backgrounds events, not shown in Fig. \ref{fig:localization}a,  that require the use of complex deconvolution algorithms. All these calculations are performed in the CDS where the Compton cones for each source can be distinguished, as exemplified in Fig. \ref{fig:localization}b.

\begin{figure}
\centering
\includegraphics[width=.9\textwidth]{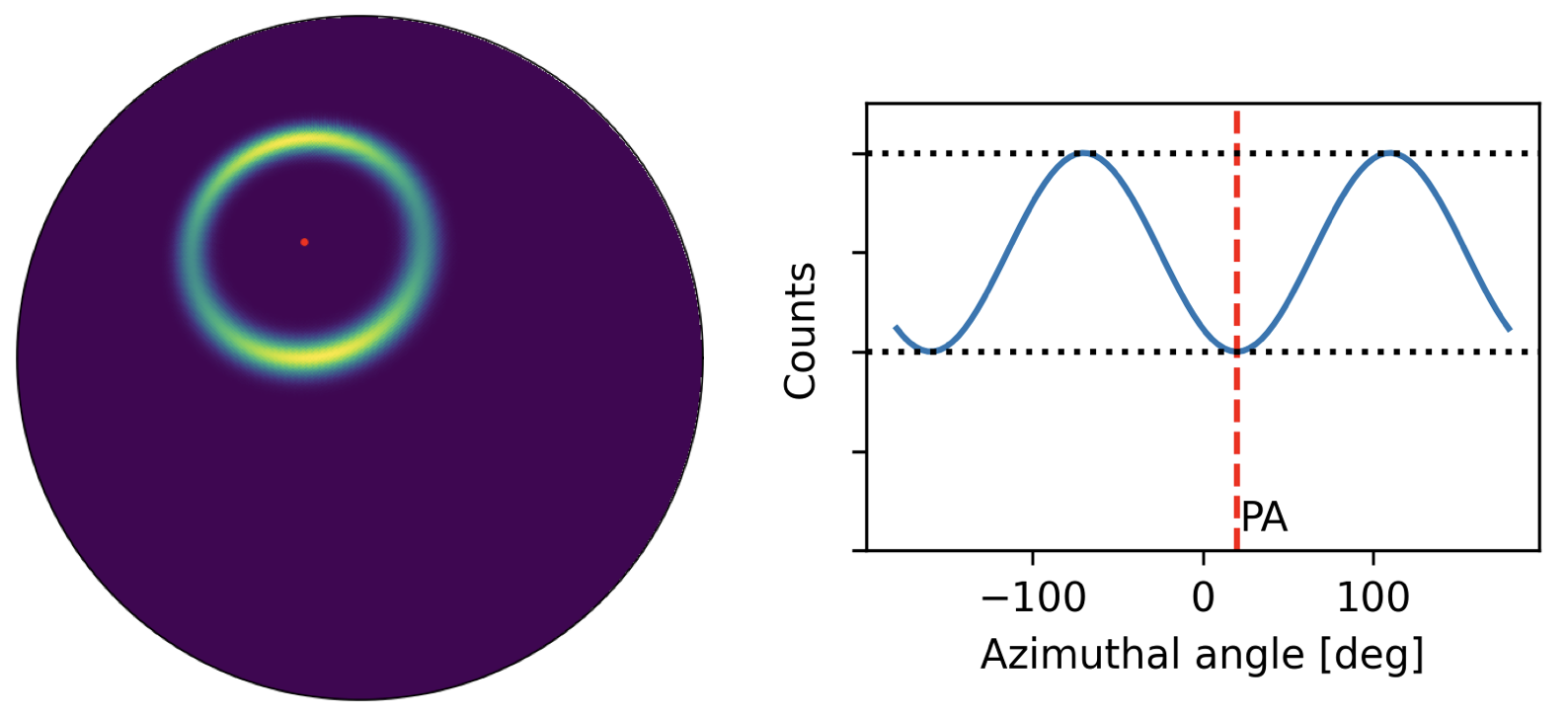}
\caption{Left: simplified point spread function (PSF) for a linearly polarized sources.  Right: projection onto the azimuth angle.  The ratio between the black dotted lines is proportional to the polarization fraction. The phase is determined by the polarization angle (PA).}
\label{fig:polarization}
\end{figure}

Polarimetry is possible thanks to the azimuthal dependence of Compton scattering in response to a linearly polarized source.  As shown in Fig. \ref{fig:polarization}, the distribution of events in the Compton cone is modulated by the polarization angle ---determining the phase--- and the polarization fraction ---proportional to the modulation factor.

\section{The cosipy library}
\label{sec:cosipy}

Although the basic principles of Compton data analysis can be applied to approximate the spectrum, location and polarization of source,  real data presents significantly more challenges.  The large number of events, high and variable backgrounds,  instrument artifacts, and the continuous rotation of the spacecraft, all call for a more complex and robust high-analysis software.  The time-dependent spectra and polarization, and correlations between energy, polarization and direction, also complicate the analysis.

\begin{figure}
\centering
\includegraphics[width=.7\textwidth]{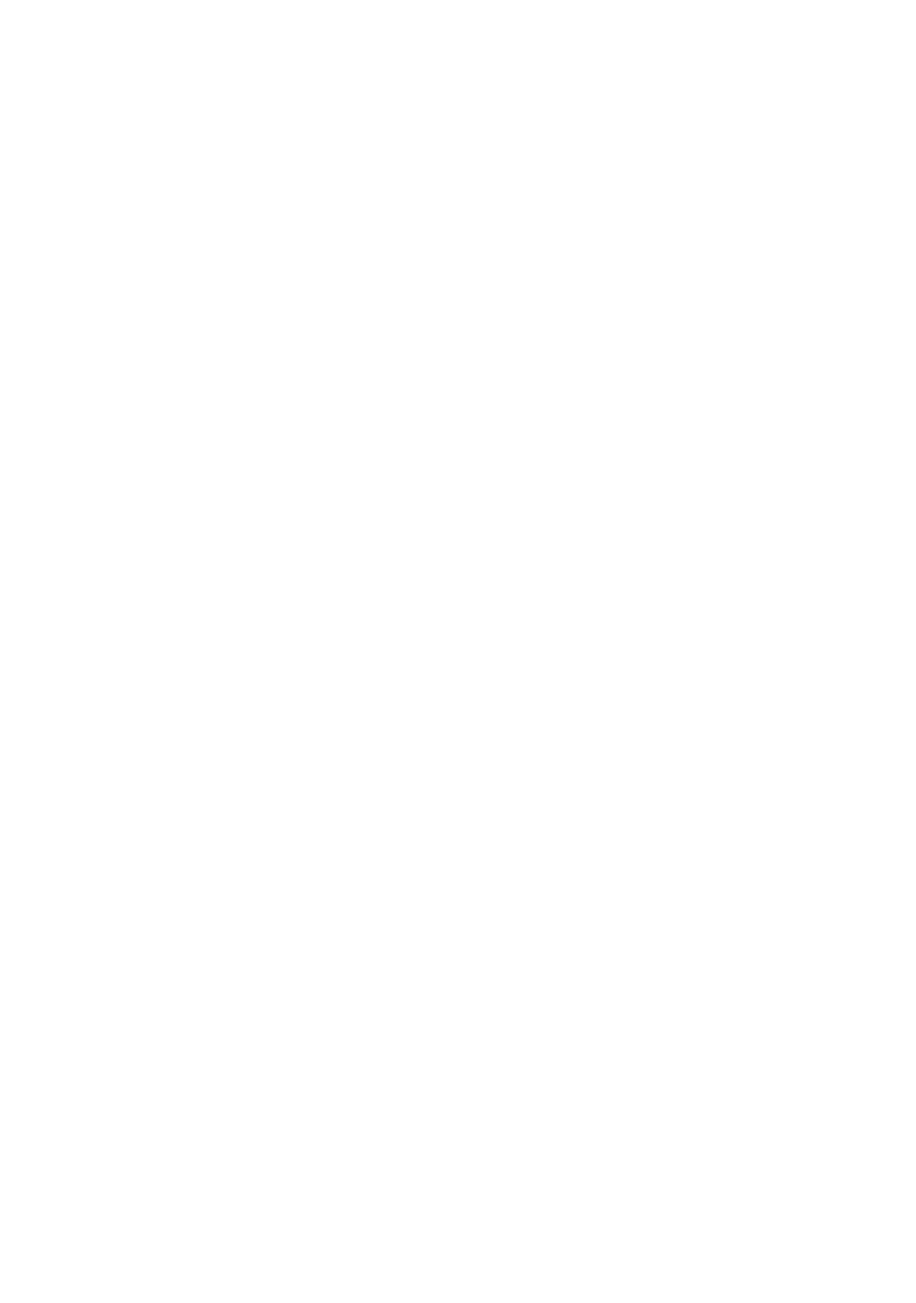}
\caption{The cosipy library will perform the  \textit{High-level Data Analysis} tasks. It is the last step of a series of software packages that compose the COSI calibration and analysis pipeline. }
\label{fig:cositools}
\end{figure}

The cosipy library is a component of a series of software packages called \textit{cositools}, and the last step of the COSI calibration and analysis pipeline, shown in \ref{fig:cositools}.  It builds upon the development during the COSI balloon campaigns.  For a full description of the calibration and pipelines see \cite{BeechertCOSICalib,2021arXiv210213158Z}.

\subsection{Current design and status}

The cosipy library follows a likelihood-based approach. It uses a forward folding technique to compare the expected measurements to the data --energy $E_m$, scattering angle $\phi$, and outgoing photon direction $(l',b')$,  which together we will call the data space $\Omega$--- given a sky model hypothesis ---i.e. a distribution of sources, each with a given spectrum.  The relationship between the expected measurements and physical photon parameters --- incoming direction $(l,b)$,  energy $E_i$ and polarization angle $PA$, which we will call the photon parameter space $\Lambda$--- is handled by the detector response,  and encoded in a a multi-dimensional matrix. 

The likelihood can be written as:

\begin{align}
\label{eq:like}
\log \mathcal{L}\lrp{\mathbf{s}, \mathbf{b}} &=  \sum_{\Delta \Omega_i} \log P\lrp{n_j|\lambda_j}  \\ 
\lambda_i \lrp{\mathbf{s}, \mathbf{b}} &= \int\displaylimits_{\Delta\Omega_i} \lrp{\int A_{\mathrm{eff}}\lrp{\Lambda}P\lrp{\Omega | \Lambda} F(\mathbf{s}, \Lambda) d\Lambda    + \mathrm{Bkg}(\mathbf{b}) }d\Omega ~, \nonumber
\end{align}

where $\mathbf{s}$ and $ \mathbf{b}$ are the signal and background parameters to fit; $P(n|\lambda)$ is the Poisson probability distribution; the term $A_{\mathrm{eff}}\lrp{\Lambda}P\lrp{\Omega|\Lambda}$ is the detector response, equal to the effective area times the probability of a detected event with physical parameters $\Lambda$ to have the measured parameters $\Omega$; and $F$ is the flux for a given location in the sky at a given energy and polarization.

The detector response is obtained from event by event Monte Carlo simulations using MEGALib \cite{2006NewAR..50..629Z}. All passive and active components are carefully modeled, and the detector effects simulated, in order to match benchmark calibration measurements. 

The cosipy library is compatible with the Multi-Mission Maximum Likelihood framework (3ML) \cite{vianello2015multi} which provides a common interface between multiple instruments. This is achieved by the use of plugins that receive a source model in a common format, and output the corresponding likelihood. This allows us to gain access to the maximum likelihood and Bayesian algorithms provided by 3ML in order to do spectral and polarization analysis.

Imaging deconvolution can also be performed by maximizing the likelihood in Eq. \ref{eq:like} over multiple hypothetical sources in a grid fully covering the sky.  Due to the high dimensionality of the problem we use the Richardson-Lucy algorithm \cite{1974AJ.....79..745L,1972JOSA...62...55R} which consists of an interactive method that converges to the maximum likelihood solution.

\begin{figure}
\centering
\includegraphics[width=.99\textwidth]{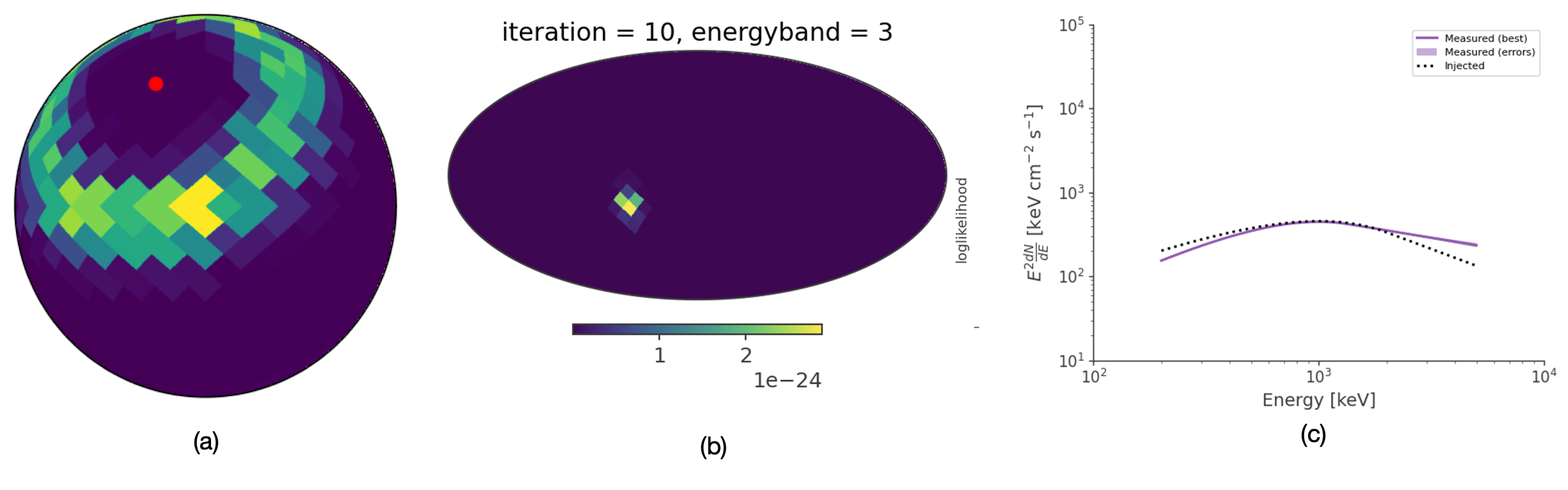}
\caption{Results from an early version of cosipy. (a) Simulated binned data, using a coarse grid, showing the location of a source (red dot) and the Compton circle.  (b) Deconvolved image of the source using the  Richardson-Lucy algorithm. (c) Fitted spectrum of the source using the 3ML framework.}
\label{fig:miniDC}
\end{figure}

We have preliminary versions of the detectors response handling, likelihood calculation, imaging and spectral analysis. Selected results are shown in Fig. \ref{fig:miniDC}. During these early stages of development we are using a simplified detector response and coarse binning in order to test and profile the code.  Currently we are working on optimizing the CPU and memory usage such that the computation can scale well to requirements of the flight version.  

\subsection{Data challenges and future plans}
\label{sec:dc-future}

We will progressively add new capabilities and improvements to cosipy,  leading up to the launch of COSI in 2027.  Starting in 2024, we will organize public data challenges with a yearly cadence, consisting of simulated data sets for sources and science topics of interest.  This will build upon the COSI balloon data challenge released in 2023 \cite{2022HEAD...1910830K}. We want to obtain feedback from the community to find issues and identify missing features required to satisfy all COSI science goals as well as to expand its portfolio.

Some of the plans for the upcoming releases include:
\begin{itemize}
\item Simultaneous location, spectral and polarization fits.
\item Time-dependent polarization analysis.
\item Implementation of alternative imaging deconvolution algorithms, such as Multiresolution Regularized Expectation Maximization,  and machine-learning-based image creation.
\end{itemize}

\section{Conclusion}

The cosipy library will perform all high-level analysis tasks to satisfy the scientific goals of the COSI mission,  an MeV gamma-ray telescope launching in 2027.  Currently under development, cosipy follows a modern likelihood-based approach, is open source, and compatible with the Multi-Mission Maximum Likelihood framework (3ML). Starting in 2024, we will make yearly releases accompanied by a public data challenge in order to obtain feedback from the community.

\bibliographystyle{JHEP} 
\bibliography{refs.bib}

%
%
%

\end{document}